\documentclass[12pt,preprint]{aastex}

\newcommand{\be}{\begin{equation}}
\newcommand{\ee}{\end{equation}}
\newcommand{\mearth}{{ M_\earth }}

\newcommand{\atime}{{ \tau_{a} }} 
\newcommand{\etime}{{ \tau_{e} }} 
\def\lta{\,\raise 0.3 ex\hbox{$ < $}\kern -0.75 em
 \lower 0.7 ex\hbox{$\sim$}\,}
\def\gta{\,\raise 0.3 ex\hbox{$ > $}\kern -0.75 em
 \lower 0.7 ex\hbox{$\sim$}\,} 

\begin{document} 

\title{\bf ACCRETION OF ROCKY PLANETS BY HOT JUPITERS} 

\author{Jacob A. Ketchum,$^{1}$, Fred C. Adams,$^{1,2}$ 
and Anthony M. Bloch$^{1,3}$} 

\affil{$^1$Michigan Center for Theoretical Physics \\
Physics Department, University of Michigan, Ann Arbor, MI 48109} 

\affil{$^2$Astronomy Department, University of Michigan, Ann Arbor, MI 48109} 

\affil{$^3$Department of Mathematics, University of Michigan, 
Ann Arbor, MI 48109} 

\begin{abstract} 

The observed population of Hot Jupiters displays a stunning variety of
physical properties, including a wide range of densities and core
sizes for a given planetary mass.  Motivated by the observational
sample, this paper studies the accretion of rocky planets by Hot
Jupiters, after the Jovian planets have finished their principal
migration epoch and become parked in $\sim4$-day orbits. In this
scenario, rocky planets form later and then migrate inward due to
torques from the remaining circumstellar disk, which also damps the
orbital eccentricity. This mechanism thus represents one possible
channel for increasing the core masses and metallicities of Hot
Jupiters. This paper determines probabilities for the possible end
states for the rocky planet: collisions with the Jovian planets,
accretion onto the star, ejection from the system, and long-term
survival of both planets. These probabilities depend on the mass of
the Jovian planet and its starting orbital eccentricity, as well as
the eccentricity damping rate for the rocky planet. Since these
systems are highly chaotic, a large ensemble ($N\sim10^3$) of
simulations with effectively equivalent starting conditions is
required.  Planetary collisions are common when the eccentricity
damping rate is sufficiently low, but are rare otherwise. For systems
that experience planetary collisions, this work determines the
distributions of impact velocities -- both speeds and impact
parameters -- for the collisions.  These velocity distributions help
determine the consequences of the impacts, e.g., where energy and
heavy elements are deposited within the giant planets.

\end{abstract} 


\section{Introduction} 

With hundreds of alien worlds detected, extrasolar planets have
dramatically completed their migration into the main-stream of
astronomy. The initial discoveries (Mayor \& Queloz 1995; Marcy \&
Butler 1996) showed that the orbital elements of extrasolar planets
are significantly different from those of Solar System planets. Some
giant planets are found in short-period orbits ($P_{orb}\approx4$
days; semi-major axes $a\approx0.05$ AU), while others have longer
orbits with a range of eccentricity, $0\le{e}\le0.9$.  Subsequent
discoveries indicate that such planetary systems are common and
display a rich variety of architectures (Marcy \& Butler 2000; Hatzes
et al. 2000; Perryman 2000, Udry et al. 2007).  The galactic planetary
census is growing rapidly, and we can probe their physical properties,
dynamics, composition, and even their weather.

An important subset of migrating Jovian planets reach the inner edge
of their parental disks, where they enter orbits with periods
$P_{orb}\sim2-5$ days.  Much of our knowledge regarding the physical
properties of extrasolar planets comes from this population, primarily
those planets observed in transit. Observations of these transiting
planets have driven an exploration of the planetary mass-radius
relation, which shows several unexpected features. The mass
distribution of these planets is wide, spanning more than three
decades. The distribution of inferred densities ranges over two orders
of magnitude, with $\rho\approx{0.16}-{26}$ g/cm$^3$.  Extrasolar
planets thus span a wide range of radii for a given mass. The
mass-radius relation for Hot Jupiters depends on many factors,
including metallicity, core mass, stellar irradiation, and additional
heat sources (Bodenheimer et al. 2003, hereafter BLL; Laughlin et
al. 2011).

This paper explores one channel for Jovian planets to change their
structure after reaching the stellar vicinity: Hot Jupiters can
accrete additional rocky bodies while they are parked in close
orbits. This accretion process increases the planetary mass, core
mass, metallicity, and density of Jovian target. This scenario works
as follows: Jovian planets stop their inward migration at semi-major
axes corresponding to $\sim4$-day orbital periods. Although the reason
for planets halting their migration is not completely understood, this
orbital radius coincides (Lin et al. 1996) with the inner truncation
point of the disk due to magnetic effects (Shu et al. 1994). As a
result, Hot Jupiters generally enter $\sim4$-day orbits with
circumstellar disk material remaining outside. Additional bodies
(rocky Earth-like planets and/or larger Neptune-like planets) can
subsequently migrate into the vicinity, where they tend to lock into
mean motion resonance with the Hot Jupiter. Although the disk acts
only on the outer rocky body, both planets continue to migrate and
interact.

The inward migration of these additional bodies, while the Hot Jupiter
is stranded inside the inner disk edge, presents an interesting
dynamical problem.  Many outcomes are possible, including collisions
between the planets, producing Earth-Jupiter systems in mean motion
resonance, and accretion of planets onto the star.  The relative
frequency of these outcomes is studied here.  If the resonant system
survives, it becomes a candidate for observing transit timing
variations (Agol et al. 2005, hereafter ASSC). If rocky planets are
accreted by the Hot Jupiter, its mass would increase. Since these
rocky bodies have higher metallicities, and densities, than the
original object, the planetary density generally increases. This
mechanism thus alters the mass-radius relationship for Hot Jupiters
and can help explain the diversity of planetary properties in the
current sample.  In particular, if the rocky bodies are large enough,
they can survive the impact (Anic et al. 2007) and increase the core
mass of the Jovian planet. The observed planet HD149026b is inferred
to have an exceptionally large core mass $M_C\sim{80}\mearth$ (Ikoma
et al. 2006, Fortney et al. 2006) and may provide one example of this
mechanism in action.

Working within this scenario, this paper shows that a large fraction
of inward migrating rocky planets collide with the Jovian planet,
thereby allowing increases in core masses and metallicities. However,
the collision rate decreases sharply for sufficiently high levels of
eccentricity damping.  If the Jovian planet has nonzero eccentricity,
and/or smaller mass, the collision rate is lower for small damping
rates, but persists for larger damping rates.  For systems that
experience planetary collisions, we determine the distributions of
impact velocities.

This paper focuses on collisions between rocky bodies and Hot
Jupiters.  A complete understanding of the planetary mass-radius
relation requires many additional mechanisms, e.g., Ohmic dissipation
in planetary atmospheres (Batygin \& Stevenson 2010, hereafter BS;
Perna et al. 2010), which are beyond the scope of this work. In
addition, Hot Jupiters display a range of spin-orbit alignments,
measured through the Rossiter-McLaughlin effect (Fabrycky \& Winn
2009); some systems may have binary companions with inclined orbits so
that planets are influenced by the Kozai effect (Wu et al.  2007).
However, this paper is limited to systems where stellar binary
companions do not play a defining role.

\section{Formulation} 

This paper studies migration scenarios where the Hot Jupiter is
already in place and a second body migrates inward. The most important
parameters are the migration rate and eccentricity damping rate for
the rocky planet, and the initial eccentricity and mass of the Jovian
planet. Given that the star and the Hot Jupiter are much more massive
than the rocky planet, the latter acts as a test particle (to leading
order).  If the rocky planet migrates sufficiently slowly, it
generally becomes locked into mean motion resonance with the Hot
Jupiter.  Continued migration of the second body then pushes both
planets inward, although this motion ceases if the second body reaches
the inner edge of the disk (and this motion becomes ineffective if the
second planet is too small).  If migration ceases, the resulting pair
of planets could survive in or near resonance. If the Hot Jupiter can
be observed in transit, the second body can produce transit timing
variations (ASSC). If migration occurs too quickly, the second planet
passes through mean motion resonance (Quillen 2006, Ketchum et al.
2011) and will often experience a close encounter with the Hot
Jupiter. The interaction event can result in either a collision
between the planets (and assimilation of the rocky body) or the
accretion of one planet (generally the smaller one) by the star.
Planets are rarely scattered out of the solar system because the
gravitational potential of the star (for a $\sim4$-day orbit) is
deeper than that of the Jovian planet (escape thus requires 3-body
effects). One goal of this work is to determine the branching ratios
for the various outcomes --- survival, accretion, scattering into the
star --- as a function of (Jovian) planetary mass and orbital
eccentricity.

We approach this problem by performing direct numerical integrations
of migrating planetary systems, i.e., we integrate the full set of 18
phase space variables for the 3-body problem consisting of the star,
Hot Jupiter, and a second migrating planet.  These integrations are
carried out using a B-S integration scheme.  In addition to gravity,
we include forcing terms that represent inward migration and
eccentricity damping; these additional effects arise due to the forces
exerted on the planet(s) by the circumstellar disk. However, we do not
model the disk directly, but rather include forcing terms to model its
behavior. 

We consider simple disk models where the surface density and
temperature distribution are power-laws in radius, 
\be\Sigma(r)=\Sigma_1\left({r_1\over{r}}\right)^p 
\qquad{\rm and}\qquad{T}(r)=T_1\left({r_1\over{r}}\right)^q,\label{powerlaw}\ee
where $\Sigma_1$ and $T_1$ are normalization constants. Here we take
$r_1=1$ AU, so the coefficients $\Sigma_1$ and $T_1$ correspond to
values at 1 AU. The index $p=1-2$, where the intermediate value
$p=3/2$ arises for the Minimum Mass Solar Nebula (Weidenschilling
1977) and where recent observations suggest $p=0.9\pm0.2$ (Andrews et
al. 2010).  The normalization for the surface density has a range of
values, with $\Sigma_1\approx1500-4500$ g/cm$^2$ (Kuchner 2004). The
power-law index of the temperature profile $q\approx3/4$ for a viscous
accretion disk (Pringle 1981) and a flat reprocessing disk (Adams \&
Shu 1986), whereas $q\approx1/2$ for a flared reprocessing disk
(Chiang \& Goldreich 1997). The latter value is often used to describe
the early solar nebula (Weidenschilling 1977).

The disk scale height $H=a_S/\Omega$, where $a_S$ is the sound speed,
which is determined by the disk temperature profile.  For a power-law
temperature distribution, we obtain the form
\be{H\over{r}}=\left({H_1\over{r_1}}\right)\left({r\over{r_1}}\right)^{(1-q)/2},\label{hover}\ee
where the scale height $H\approx{0.1}{r}$ at $r_1=1$ AU. 

To account for planet migration, we assume that the semi-major axis of
the outer planet decreases with time according to the ansatz
\be\dot{a}/a=-1/\atime,\label{adamp}\ee 
where $\atime$ is the migration timescale, which varies with $a$.  
We assume that only the outer planet experiences torques from the
circumstellar disk.  Small planets (less massive than Saturn) cannot
clear disk gaps, and migrate inward quickly through the process of
Type~I migration (Ward 1997).  Larger bodies clear gaps and migrate
more slowly. Planets are expected to experience a range of migration
rates, depending on planet masses and disk properties.  Estimates of
the migration timescale for $a\sim{1}$ AU typically fall in the range
$10^4-10^5$ yr (Goldreich \& Tremaine 1980, Papaloizou \& Larwood
2000).  The migration timescale decreases with semi-major axis $a$ and
can be modified by subkeplerian rotation (Adams et al. 2009). Since we
must perform a large ensemble of simulations using effectively
equivalent starting conditions, we adopt a relatively simple model of
Type~I migration. 
 
The strength of Type~I torques can be written in the form 
\be{T_I}=f_1\left({m_P\over{M_\ast}}\right)^2 
\pi\Sigma{r^2}(r\Omega)^2\left({r\over{H}}\right)^2,\label{torqueone}\ee
where $m_P$ is the mass of the rocky planet and $f_1\approx0.6$ is a 
dimensionless parameter (Ward 1997, Tanaka et al. 2002). For nearly 
Keplerian disks, the orbital angular momentum for a circular orbit is
given by $J=m_P(GM_\ast{r})^{1/2}$, and the migration timescale $\atime$ becomes 
\be\atime={J\over{T_I}}={1\over{f_1}}\left({M_\ast\over{m_P}}\right)
\left({M_\ast\over\pi\Sigma{r^2}}\right)\left({H\over{r}}\right)^2{1\over\Omega}.\label{timeone}\ee
Using typical parameter values, we obtain the scaled result 
\be\atime=5.6\times10^4\,\,{\rm yr}\,\,
\left({r\over{r_1}}\right)^{p-q+1/2}\left({m_P\over10\mearth}\right)^{-1}.\ee 
We adopt the indices used to model the early solar nebula, $p=3/2$ and $q=1/2$, 
so the migration timescale is proportional to the orbital period, 
\be\atime\approx{56,000}\,P_{orb}\,(m_P/10\mearth)^{-1},\label{awork}\ee
where the period is in years. The timescale $\atime$ thus decreases as
the rocky planet moves inward, i.e., migration accelerates.

In addition to inward migration, circumstellar disks damp the 
orbital eccentricity $e$ of the migrating planet. This damping effect
arises in almost all numerical simulations of the process (e.g., 
Kley et al. 2004), and can be parameterized through the ansatz 
\be\dot{e}/e=-1/\etime=K(\dot{a}/a)\qquad{\rm so}\,\,{\rm that}\qquad\etime=\atime/K,\label{edamp}\ee
where $\etime$ is the eccentricity damping timescale. For planets that
are large enough to clear gaps, analytic calculations suggest that
eccentricity can be excited through the action of disk torques
(Goldreich \& Sari 2003, Ogilvie \& Lubow 2003), although multiple
planet systems would be compromised if this were always the case
(Moorhead \& Adams 2005). For smaller planets that remain embedded
only eccentricity damping is expected. Given the uncertainties, we
parameterize the eccentricity damping using equation (\ref{edamp}) and
explore a wide range of the damping parameter $K$ such that
$10^{-2}\le{K}\le10^2$, where fully embedded planets are expected to
have $K$ values at the high end of this range (Artymowicz 1993).

Note that this treatment implicitly assumes that the migrating planets
are small enough so that they produce no back reaction on the disk. 
Since we are primarily interested in planetary cores in the mass
range $m_P=1-30\mearth$, this assumption is expected to be valid.

\section{Results} 

Using the formulation outlined above, we study the inward migration of
rocky planets in planetary systems that contain a Hot Jupiter. The
primary objective is to catalog the probabilities of the various
outcomes, including survival, collisions, and accretion onto the
star. A secondary goal is to determine the distribution of impact
velocities for those cases that end in planetary collisions.

The parameter space for this study is large. For the sake of
definiteness, the star has mass $M_\ast={1.0}M_\odot$ and the Jovian
planet has starting semi-major axis $a=0.05$ AU ($P_{orb}\approx{4}$
day). The eccentricity of the giant planet varies over the range
${0}\le{e}\le{0.3}$ (these planets are expected to become tidally
circularized, but only on much longer timescales). The rocky planet
starts just outside the 5:1 mean motion resonance ($a\approx0.15$ AU),
with small eccentricity $e=0.001$, and fixed mass $m_P=10\mearth$; in
this problem, the rocky planet acts like a test particle, so its mass
cannot greatly affect the dynamics. The migration rate of the rocky
planet varies with location, according to equation (\ref{awork});
inside the disk edge ($a\lta{0.05}$AU), migration ceases. With these
specifications, we consider the effects of varying the mass and
eccentricity of the Jovian planet, and the eccentricity damping rate
(through $K$) of the rocky planet. Since these systems are highly
chaotic, a large ensemble of numerical experiments must be performed
for each point in parameter space ($\sim$1000 independent
realizations).

\begin{figure} 
\figurenum{1} 
{\centerline{\epsscale{0.90} \plotone{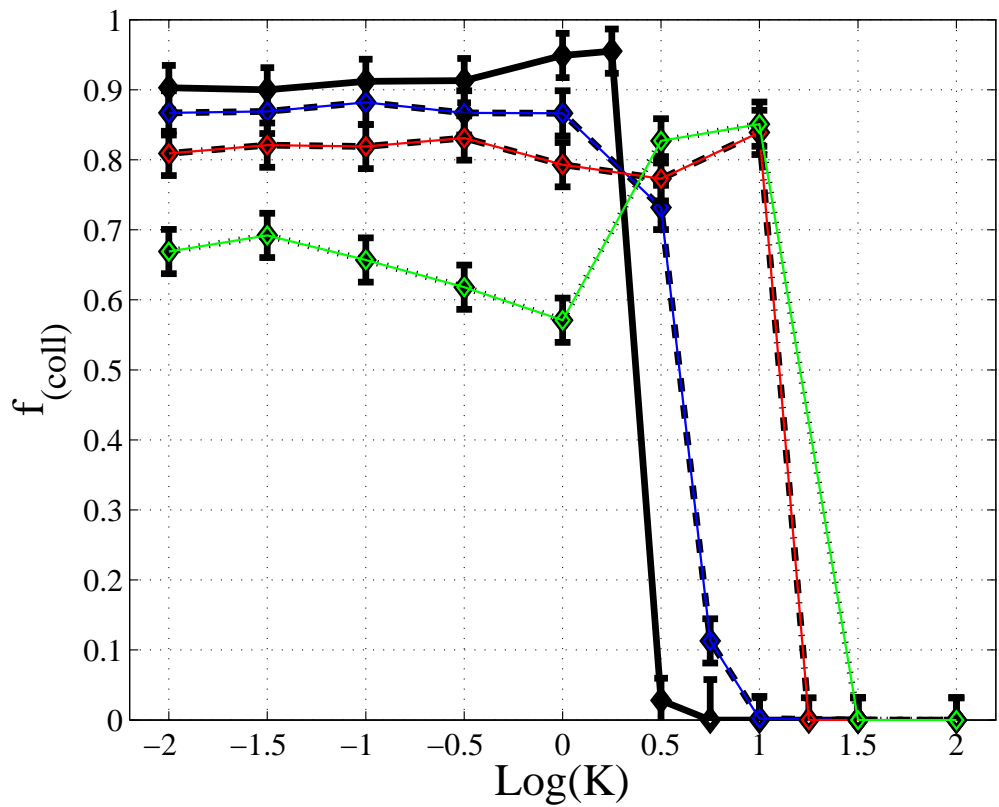} } } 
\figcaption{Collision fraction for rocky planets impacting Hot 
Jupiters versus eccentricity damping parameter $K$. The curves 
correspond to varying initial eccentricity of the Jovian orbit: 
$e=0$ (black-solid), $e=0.1$ (blue-dashes), $e=0.2$ (red-dot-dashes),
and $e=0.3$ (green-dots). }
\label{fig:efraction} 
\end{figure}

\begin{figure} 
\figurenum{2} 
{\centerline{\epsscale{0.90} \plotone{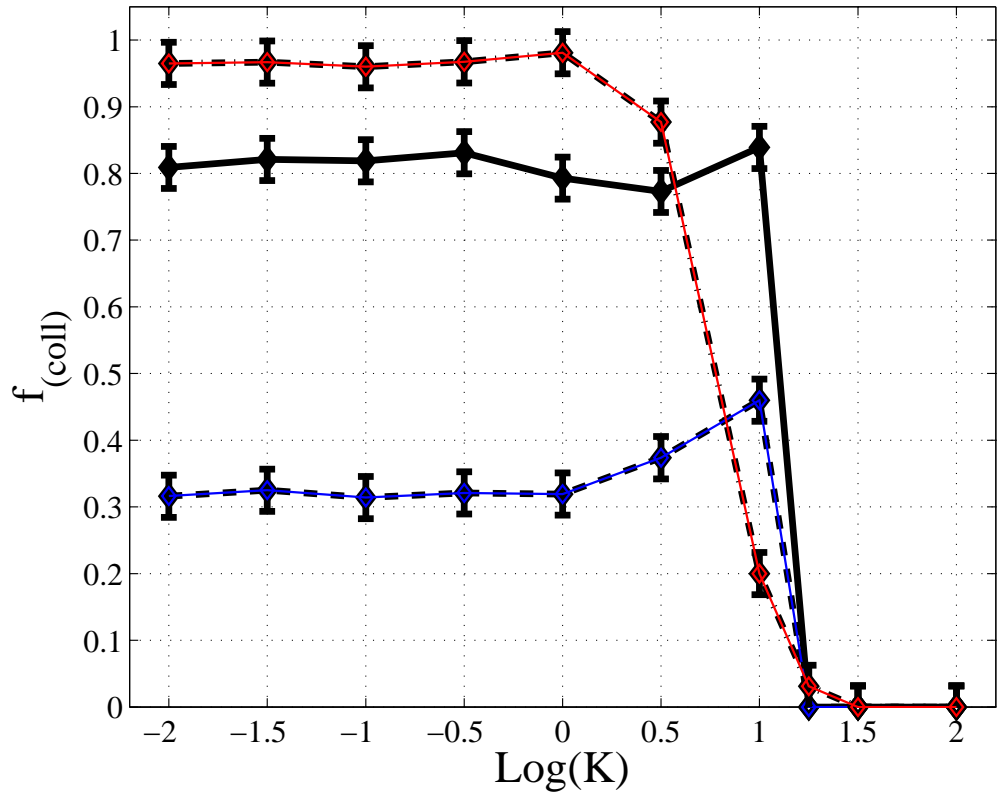} } } 
\figcaption{Collision fraction for rocky planets impacting Hot 
Jupiters versus eccentricity damping parameter $K$. The curves
correspond to varying masses of the Jovian planet: $0.5M_J$
(red-dot-dashes-top), $1M_J$ (black-solid-middle), and $2M_J$
(blue-dashes-bottom).}
\label{fig:mfraction} 
\end{figure}

The main result from these simulations is the fraction of the trials
that end with the two planets colliding. For a given migration rate,
collisions represent the most common outcome provided that
eccentricity damping is not too effective. These results are depicted
in Figures \ref{fig:efraction} and \ref{fig:mfraction}, which show the
fraction of collisions plotted versus the parameter $K$ that sets the
strength of eccentricity damping for the rocky planet (equation
[\ref{edamp}]).  Figure \ref{fig:efraction} shows collision fractions
for four choices of starting eccentricity for the Hot Jupiter, from
$e=0$ to $e=0.3$. Figure \ref{fig:mfraction} shows collision fractions
for fixed starting eccentricity $e=0.2$ and three choices for the Hot
Jupiter mass, $M_P/M_J=$ 0.5, 1, and 2. In both Figures, each point
shown corresponds to the fractions calculated from $N\sim{1000}$
independent realizations of the starting conditions. The error bars
($\sim1/\sqrt{N}$)  provide a crude measure of the uncertainties. 

\begin{figure} 
\figurenum{3} 
{\centerline{\epsscale{0.75} \plotone{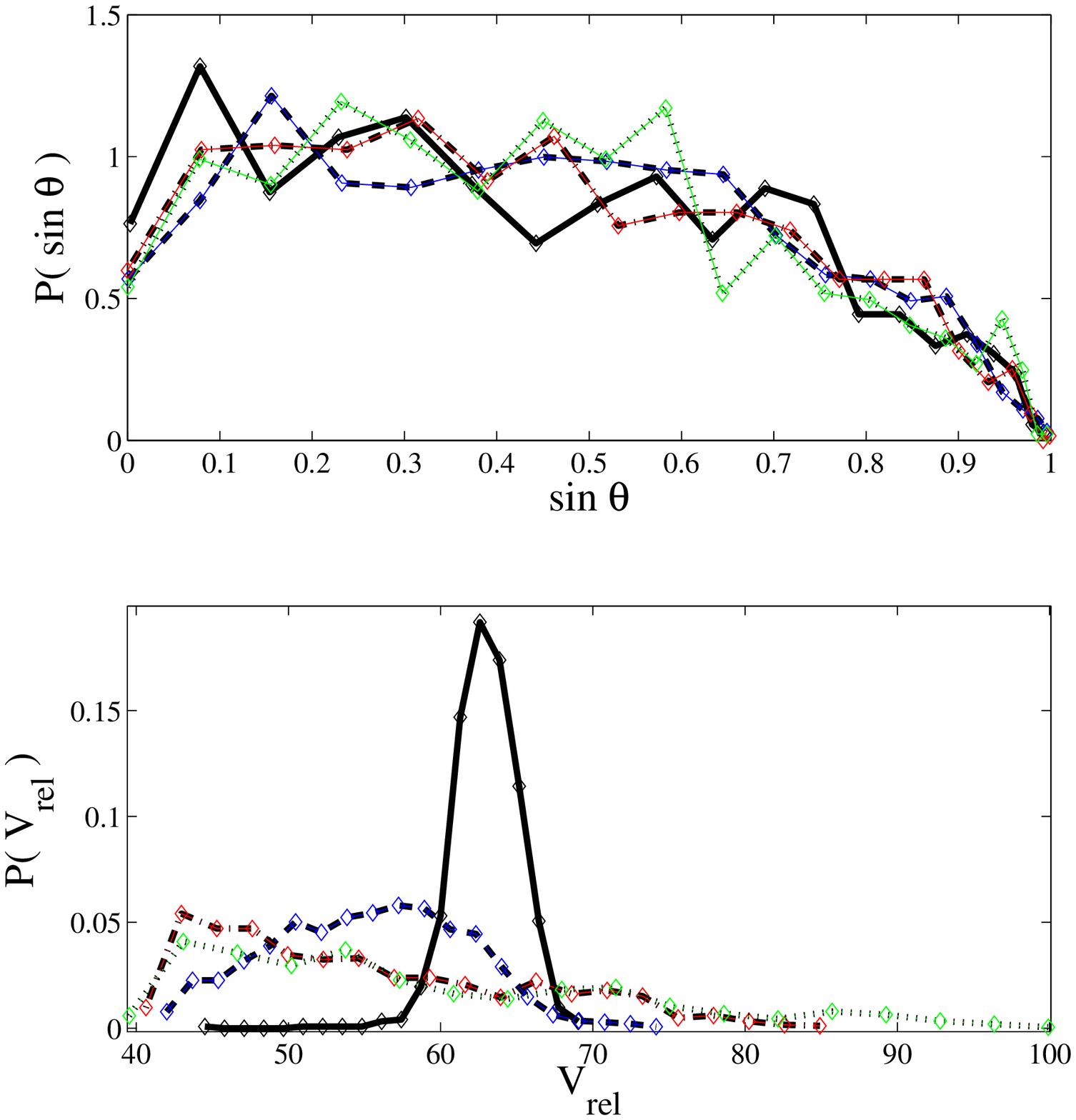} } } 
\figcaption{Distribution of impact velocities (for $K=1$). Top 
panel shows distributions of impact directions, specified by
$\sin\theta$, where the angle $\theta$ is given by
$\cos\theta=\hat{v}\cdot\hat{r}$. Bottom panel shows distributions of
impact speeds. In both panels, distributions are shown for systems
where the Jovian planet has initial eccentricity $e=0$ (black-solid),
$e=0.1$ (blue-dashes), $e=0.2$ (red-dot-dashes), and $e=0.3$
(green-dots). }
\label{fig:vdist} 
\end{figure}

The results displayed in Figures \ref{fig:efraction} and
\ref{fig:mfraction} show a robust trend: For sufficiently weak
eccentricity damping, $K<K_C\approx{10}$, most simulations end with
collisions between the planets.  For stronger eccentricity damping,
$K>K_C$, the collision fraction becomes negligible and nearly all of
the systems survive (keeping both planets) over the entire range of
integration times. Further, the critical level of eccentricity damping
($K_C$) depends on the starting eccentricity and mass of the Hot
Jupiter.

Larger eccentricities (for the Jovian orbit) allow collisions to occur
in the face of greater eccentricity damping, following a trend of the
approximate form $\log_{10}{K_C}\approx(3+10e)/4$ (from fitting).
However, larger eccentricities combined with smaller $K$ values yield
lower collision rates. In this regime, collision events are replaced
(primarily) by accretion events (onto the star). The larger
eccentricity of the Jovian planet provides the rocky planet with
greater opportunity to pass by and enter the gravitational realm of
the star. Similarly, larger masses for the Jovian planet allow
collisions to occur for larger values of the eccentricity damping
parameter. In addition, larger masses combined with smaller $K$ values
lead to lower collision rates. In this case, the collision events are
(again) replaced with accretion events. The larger mass of the Jovian
planet can scatter the rocky planet before impact, and the scattering
alters the orbit of the rocky planet enough to send it into the star
(or, more rarely, eject the planet).

These results were obtained for a single migration rate; for faster
(slower) migration, the outer planet is less (more) likely to lock
into mean motion resonance and is more (less) likely to collide with
the Jovian planet (Ketchum et al. 2011).  We have performed additional
simulations with faster migration (not shown) to confirm these trends.

For simulations that end in planetary collisions, the effect on the
Jovian planet depends on the impact velocity of the rocky planet.
Distributions of these impact velocities are depicted in Figure
\ref{fig:vdist} (for eccentricity damping parameter $K=1$). The top
panel shows distributions of the angle at which the incoming planet
strikes the giant planet surface. This distribution is equivalent to
the distribution of impact parameter $\varpi=R_P\sin\theta$.
Collision dynamics depend on the impact speed $v_{rel}$, shown in the
bottom panel of Figure \ref{fig:vdist}, and the escape speed
$v_{esc}=(GM_P/R_P)^{1/2}\approx37$ km/s (for $M_P=1M_J$ and
$R_P=1.4R_J$). In the limit $v_{rel}\gg{v_{esc}}$, the giant planet
presents a circular target and the probability
$P(\varpi)\propto{P}(\sin\theta)$ increases with impact parameter
$\varpi$.  In the limit $v_{rel}\ll{v_{esc}}$, gravity focuses
incoming trajectories into nearly radial paths and the distribution
peaks near $\varpi=0$. The calculated distribution is relatively flat,
but falls with $\varpi$, which suggests significant gravitational
focusing. This expectation is validated in the bottom panel, which
shows that the impact speeds fall in the range $v\sim40-100$ km/s,
i.e., $v/v_{esc}\sim1-3$. When the Jovian planet has nonzero
eccentricity, the velocity distribution shows a broad peak near $v=50$
km/s.  For systems with $e=0$, however, the distribution has a
narrower peak near $v=65$ km/s. One reason for this difference is that
the rocky planets migrate further inward (before colliding) when
$e=0$, so they are deeper in the gravitational potential well of the
star. Obtaining a greater dynamical understanding of this trend
provides an interesting problem for future investigation.

\section{Conclusion} 

This paper explores the accretion of rocky planetary bodies by Hot
Jupiters after they reach close-in orbits. The results show that
collisions between planets are common when the eccentricity damping
rate is sufficiently small, and rare otherwise. In approximate terms,
collisions require the eccentricity damping parameter
$K\le{K_C}\approx{10}$, where the threshold $K_C$ depends on the
eccentricity and mass of the Jovian planet (Figures
\ref{fig:efraction}, \ref{fig:mfraction}).  The corresponding
distributions of impact velocities for the collisions are shown in
Figure \ref{fig:vdist}.

These results have important implications for the diversity seen in
the observational sample of Hot Jupiters: For large $K$ values, both
planets usually survive, in resonance, and such systems can exhibit
observable transit timing variations (ASSC). For small $K$ values,
collisions are common whenever disks produce rocky bodies after a Hot
Jupiter has migrated to its inner orbit. These collisions, in turn,
can increase the core mass and the metallicity of the Jovian
planet. Accretion onto the star and ejection are almost always rare.

The frequency of collisions is governed by the $K$ value, which
depends on disk structure, viscosity, and the mass of the migrating
rocky planet. Previous studies of planet-disk interactions generally
find $K$ values of order unity for migrating planets that clear gaps
(Kley et al. 2004), but $K\approx10-30$ for smaller embedded planets
(Artymowicz 1993). The outcomes thus depend on gap-clearing. For
low-viscosity disks, planets clear gaps when their Hill sphere exceeds
the disk scale height, $r_H>H$ (Crida et al. 2006; Papaloizou \&
Terquem 2006), which requires $m_P\gta27\mearth$ for the disk
parameters used here. The gap doesn't need to be completely open to
reduce the $K$ value below the threshold $K_C$. Nonetheless, relatively
large rocky planets ($m_P\gta10-20\mearth$) are required for partial
gap-clearing, reduced $K$ values, and hence collisions.  Small planets
with $m_P\lta10\mearth$ are expected to have $K>K_C$ and hence to
avoid collision with high probability.  In addition, incoming rocky
bodies must survive the collision and reach the core to increase its
mass; survival is expected when $m_P\gta{1-10}\mearth$ (Anic et
al. 2007). Both the occurrence of collisions and subsequent survival
to reach the core thus require $m_P\gta10\mearth$. Although this
threshold mass should be determined more rigorously, these results
show that larger rocky planets have more influence (per unit mass)
than smaller ones.

In addition to increasing the core mass, accretion of rocky planets
can affect the energy budget of giant planets. Figure \ref{fig:vdist}
shows the distribution of impact speeds for rocky planets that collide
with Hot Jupiters. This distribution indicates speeds $v\sim40-100$
km/s, so we consider a benchmark $v\sim60$ km/s. With this speed, an
accreting ``superearth'' planet with mass $m_P=10\mearth$ deposits
energy
\be\Delta{E}={1\over2}m_P{v^2}\approx1.1\times10^{42}{\rm ergs}.\ee 
To put this energy increment into perspective, note that the binding
energy of the Hot Jupiter $U=fGM_P^2/R_P\approx1.6\times10^{43}$ erg
(using typical values $M_P=1M_J$, $R_P=1.4M_J$, and $f=3/5$). A single
collision thus accounts for $\sim7\%$ of the binding energy of a Hot
Jupiter. If we assume the energy $\Delta{E}$ is deposited deep within
the planet, and slowly leaks out over time $\Delta{t}\sim1$ Gyr, the
associated power increment $\Delta{P}\approx3.5\times10^{18}$ W, large
enough to help inflate the planetary radius (BLL,BS). On the other
hand, if the energy is deposited in the upper atmosphere of the
planet, it quickly radiates away and cannot inflate the radius.

The results of this paper pose a number of interesting problems for
future work.  To determine the number of accretion events (per Hot
Jupiter) we need a better understanding of eccentricity damping rates
for both migrating rocky planets and Hot Jupiters; we also need
estimates for the number (and masses) of rocky planets produced after
Hot Jupiter migration has occurred. When accretion events take place,
we need to understand the energy deposition within the giant planet
and the subsequent long-term transfer of heat/energy out of the
planetary body. These issues, and others, will help explain the
observed diversity in the properties of Hot Jupiters.

\newpage

\acknowledgments

This work was supported by NSF grant DMS-0806756 from the Division of
Applied Mathematics, NASA grant NNX11AK87G (FCA), and NSF grant
DMS-0907949 (AMB).

\end{document}